\documentclass[twocolumn,tighten]{aastex63}
\usepackage{amsmath}
\usepackage{threeparttable}
\usepackage{float}
\usepackage{graphicx}
\usepackage{CJK}
\usepackage{xspace}

\newcommand{\package}[1]{\tt{#1}}

\shorttitle{Luminous Galaxies at $z\approx10-12$}
\shortauthors{Naidu \& Oesch et al.}
\begin{document}
\begin{CJK*}{UTF8}{gbsn}

\title{Two Remarkably Luminous Galaxy Candidates at $z\approx10-12$ Revealed by \textit{JWST}}

\correspondingauthor{Rohan P. Naidu, Pascal A. Oesch}
\email{rnaidu@mit.edu, pascal.oesch@unige.ch}
\author[0000-0003-3997-5705]{Rohan P. Naidu}
\thanks{NASA Hubble Fellow}
\affiliation{Center for Astrophysics $|$ Harvard \& Smithsonian, 60 Garden Street, Cambridge, MA 02138, USA}
\affiliation{MIT Kavli Institute for Astrophysics and Space Research, 77 Massachusetts Ave., Cambridge, MA 02139, USA}
\author[0000-0001-5851-6649]{Pascal A. Oesch}
\affiliation{Department of Astronomy, University of Geneva, Chemin Pegasi 51, 1290 Versoix, Switzerland}
\affiliation{Cosmic Dawn Center (DAWN), Niels Bohr Institute, University of Copenhagen, Jagtvej 128, K\o benhavn N, DK-2200, Denmark}
\author[0000-0002-8282-9888]{Pieter van Dokkum}
\affiliation{Astronomy Department, Yale University, 52 Hillhouse Ave, New Haven, CT 06511, USA}
\author[0000-0002-7524-374X]{Erica J. Nelson}
\affiliation{Department for Astrophysical and Planetary Science, University of Colorado, Boulder, CO 80309, USA}
\author[0000-0002-1714-1905]{Katherine A. Suess}
\affiliation{Department of Astronomy and Astrophysics, University of California, Santa Cruz, 1156 High Street, Santa Cruz, CA 95064 USA}
\affiliation{Kavli Institute for Particle Astrophysics and Cosmology and Department of Physics, Stanford University, Stanford, CA 94305, USA}
\author[0000-0003-2680-005X]{Gabriel Brammer}
\affiliation{Cosmic Dawn Center (DAWN), Niels Bohr Institute, University of Copenhagen, Jagtvej 128, K\o benhavn N, DK-2200, Denmark}
\author[0000-0001-7160-3632]{Katherine E. Whitaker}
\affil{Department of Astronomy, University of Massachusetts, Amherst, MA 01003, USA}
\affil{Cosmic Dawn Center (DAWN), Denmark}
\author[0000-0002-8096-2837]{Garth Illingworth}
\affiliation{Department of Astronomy and Astrophysics, University of California, Santa Cruz, CA 95064, USA}
\author[0000-0002-4989-2471]{Rychard Bouwens}
\affiliation{Leiden Observatory, Leiden University, NL-2300 RA Leiden, Netherlands}
\author[0000-0002-8224-4505]{Sandro Tacchella}
\affiliation{Kavli Institute for Cosmology, University of Cambridge, Madingley Road, Cambridge, CB3 0HA, UK}
\affiliation{Cavendish Laboratory, University of Cambridge, 19 JJ Thomson Avenue, Cambridge, CB3 0HE, UK}
\author[0000-0003-2871-127X]{Jorryt Matthee}
\affiliation{Department of Physics, ETH Z\"urich, Wolfgang-Pauli-Strasse 27, 8093 Z\"urich, Switzerland}
\author[0000-0001-9610-7950]{Natalie Allen}
\affiliation{Cosmic Dawn Center (DAWN), Niels Bohr Institute, University of Copenhagen, Jagtvej 128, K\o benhavn N, DK-2200, Denmark}
\author[0000-0001-5063-8254]{Rachel Bezanson}
\affiliation{Department of Physics and Astronomy and PITT PACC, University of Pittsburgh, Pittsburgh, PA 15260, USA}
\author[0000-0002-1590-8551]{Charlie Conroy}
\affiliation{Center for Astrophysics $|$ Harvard \& Smithsonian, 60 Garden Street, Cambridge, MA 02138, USA}
\author[0000-0002-2057-5376]{Ivo Labbe} 
\affiliation{Centre for Astrophysics and Supercomputing, Swinburne University of Technology, Melbourne, VIC 3122, Australia}
\author[0000-0001-6755-1315]{Joel Leja}
\affil{Department of Astronomy \& Astrophysics, The Pennsylvania State University, University Park, PA 16802, USA}
\affil{Institute for Computational \& Data Sciences, The Pennsylvania State University, University Park, PA, USA}
\affil{Institute for Gravitation and the Cosmos, The Pennsylvania State University, University Park, PA 16802, USA}
\author[0000-0002-5757-4334]{Ecaterina Leonova}
\affiliation{GRAPPA, Anton Pannekoek Institute for Astronomy and Institute of High-Energy Physics, University of Amsterdam, Science Park 904, 1098 XH Amsterdam, The Netherlands}
\author[0000-0002-6668-2011]{Dan Magee}
\affiliation{UCO/Lick Observatory, University of California, Santa Cruz, CA, 95064}
\author[0000-0002-0108-4176]{Sedona H. Price}
\affiliation{Max-Planck-Institut f\"{u}r extraterrestrische Physik (MPE), Giessenbachstr. 1, D-85748 Garching, Germany}
\author[0000-0003-4075-7393]{David J. Setton}
\affiliation{Department of Physics and Astronomy and PITT PACC, University of Pittsburgh, Pittsburgh, PA 15260, USA}
\author[0000-0002-6338-7295]{Victoria Strait}
\affiliation{Cosmic Dawn Center (DAWN), Niels Bohr Institute, University of Copenhagen, Jagtvej 128, K\o benhavn N, DK-2200, Denmark}
\author[0000-0001-7768-5309]{Mauro Stefanon}
\affil{Departament d'Astronomia i Astrof\`isica, Universitat de Val\`encia, C. Dr. Moliner 50, E-46100 Burjassot, Val\`encia,  Spain}
\affil{Unidad Asociada CSIC "Grupo de Astrof\'isica Extragal\'actica y Cosmolog\'ia" (Instituto de F\'isica de Cantabria - Universitat de Val\`encia)}
\author[0000-0003-3631-7176]{Sune Toft}
\affiliation{Cosmic Dawn Center (DAWN), Niels Bohr Institute, University of Copenhagen, Jagtvej 128, K\o benhavn N, DK-2200, Denmark}
\author[0000-0003-1614-196X]{John R. Weaver}
\affiliation{Department of Astronomy, University of Massachusetts, Amherst, MA 01003, USA}
\author[0000-0001-8928-4465]{Andrea Weibel}
\affiliation{Department of Astronomy, University of Geneva, Chemin Pegasi 51, 1290 Versoix, Switzerland}

\begin{abstract}
The first few hundred Myrs at $z>10$ mark the last major uncharted epoch in the history of the Universe, where only a single galaxy (GNz11 at $z\approx11$) is currently spectroscopically confirmed. Here we present a search for luminous $z>10$ galaxies with \textit{JWST}/NIRCam photometry spanning $\approx1-5\mu$m and covering 49 arcmin$^{2}$ from the public \textit{JWST} Early Release Science programs (CEERS and GLASS). Our most secure candidates are two $M_{\rm{UV}}\approx-21$ systems: GLASS-z12 and GLASS-z10. These galaxies display abrupt $\gtrsim1.8$ mag breaks in their spectral energy distributions, consistent with complete absorption of flux bluewards of Lyman-$\alpha$ that is redshifted to $z=12.4^{+0.1}_{-0.3}$ and $z=10.4^{+0.4}_{-0.5}$. Lower redshift interlopers such as quiescent galaxies with strong Balmer breaks would be comfortably detected at $>5\sigma$ in multiple bands where instead we find no flux. From SED modeling we infer that these galaxies have already built up $\sim 10^9$ solar masses in stars over the $\lesssim300-400$ Myrs after the Big Bang. The brightness of these sources enable morphological constraints. Tantalizingly, GLASS-z10 shows a clearly extended exponential light profile, potentially consistent with a disk galaxy of $r_{\rm{50}}\approx0.7$ kpc. These sources, if confirmed, join GNz11 in defying number density forecasts for luminous galaxies based on Schechter UV luminosity functions, which require a survey area $>10\times$ larger than we have studied here to find such luminous sources at such high redshifts. They extend evidence from lower redshifts for little or no evolution in the bright end of the UV luminosity function into the cosmic dawn epoch, with implications for just how early these galaxies began forming.  This, in turn, suggests that future deep \textit{JWST} observations may identify relatively bright galaxies to much earlier epochs than might have been anticipated.
\end{abstract}


\section{Introduction}
\label{sec:introduction}

When and how the first galaxies formed remains one of the most intriguing questions of extragalactic astronomy and observational cosmology \citep[see][for recent reviews]{DayalReview18,Robertson21}. Although deep observations with the Hubble Space Telescope (HST) have pushed our cosmic horizon to within the first 400 Myr of the Big Bang, galaxies at $z\gtrsim 12$ cannot be observed with HST due to the limit of its wavelength coverage at 1.6 \micron.

With the advent of JWST, we now have an unprecedented view of the Universe at $\sim2-5$ \micron\ thanks to the extremely sensitive NIRCam instrument \citep[see, e.g.,][]{RiekeNircam}. The extended wavelength coverage enables the study of rest-frame optical wavelengths up to $z\sim10$ and allows for rest-frame UV selections of galaxies out to much higher redshifts. 

Here we present first results from a search for particularly luminous $z>10$ sources across the two JWST Early Release Science deep fields. The most luminous galaxies are of particular importance. They may trace overdensities and thus pinpoint where galaxy formation first started in the early Universe \citep[e.g.,][]{Leonova21,Endsley21,Larson22}. Furthermore, they provide the most stringent constraints on early galaxy build-up and promise rich scientific returns.

One particular example of this is provided by GN-z11 \citep{Oesch16} that was detected with HST. Its discovery in the two CANDELS/GOODS fields that cover a search volume of only $\sim10^6$ Mpc$^3$ was initially quite surprising. Theoretical and empirical models of early galaxy formation predicted that a 10-100$\times$ larger survey would have been required to find one such bright galaxy at $z=11$ \citep[e.g.][]{Waters16,Mutch16}. This highlights the potential of the brightest galaxies at the cosmic frontier to set unique constraints on the physics of galaxy formation \citep[see also][]{Behroozi18}. In particular, the number density of such bright sources, i.e. the bright end cutoff of the UV luminosity function, is a very powerful tool to test the efficiency of star-formation and potential feedback mechanisms in the very early Universe \citep[][]{Bowler14, Tacchella18,Bowler20}. 

These results have been extended over the last few years, and evidence is emerging for a differential evolution of the galaxy population during the reionization epoch at $z>6$. While the number densities of fainter galaxies continue to decline with redshift, the most UV-luminous sources seem to be in place rather early \citep[e.g.,][]{Stefanon19,Bowler20,Morishita20,Harikane22,Bagley22}. Furthermore, several authors found evidence for pronounced Balmer breaks in bright $z\sim8-10$ galaxies, which would indicate a very early formation epoch with intense star-formation \citep[e.g.,][]{Hashimoto18,Roberts-Borsani20,Laporte21}. However, others find very young ages for the average population \citep[][]{Stefanon21,Stefanon22}. These inferences, at the moment, are also highly sensitive to the prior adopted on the star-formation history \citep[e.g.,][]{Tacchella22,Whitler22}. Timing the onset of first star-formation in bright galaxies out to $z\sim 10$ is thus still highly uncertain, and probing the number density of the most luminous sources at even higher redshifts is the most direct way of addressing this.

The brightest galaxies are also the most amenable to follow-up studies through spectroscopy. For instance, GN-z11 had its redshift confirmed both through grism spectroscopy with HST and with emission lines from ground-based Keck observations \citep[see][]{Oesch16,Jiang21}. With the combination of NIRCam and NIRSpec, we enter a new era -- the rest-frame optical features of galaxies in the Epoch of Reionization ($z\approx6-9$, e.g., \citealt{Mason19nonpar}) will come fully into view. However, the strongest emission lines of galaxies at $z>10$ will still remain out of reach. Spectroscopically confirming such sources will require measuring order of magnitude fainter lines or strong continuum breaks. Identifying luminous systems at $z>10$ to facilitate these measurements is therefore a crucial step in fulfilling JWST's mission of charting cosmic dawn.

This paper 
is organized as follows -- \S\ref{sec:data} describes the datasets analyzed in this work, \S\ref{sec:Sample} presents our sample selection, and in section \S\ref{sec:results} we show our results. A discussion on the implications follows in \S\ref{sec:discussion}, before we conclude with a summary and an outlook in \S\ref{sec:summary}.

Magnitudes are in the AB system \citep[e.g.,][]{Oke83}. For summary statistics we report medians along with 16$^{\rm{th}}$ and 84$^{\rm{th}}$ percentiles. We adopt a \citet{Planck2015} cosmology.

\begin{figure*}
\centering
\includegraphics[width=0.95\linewidth]{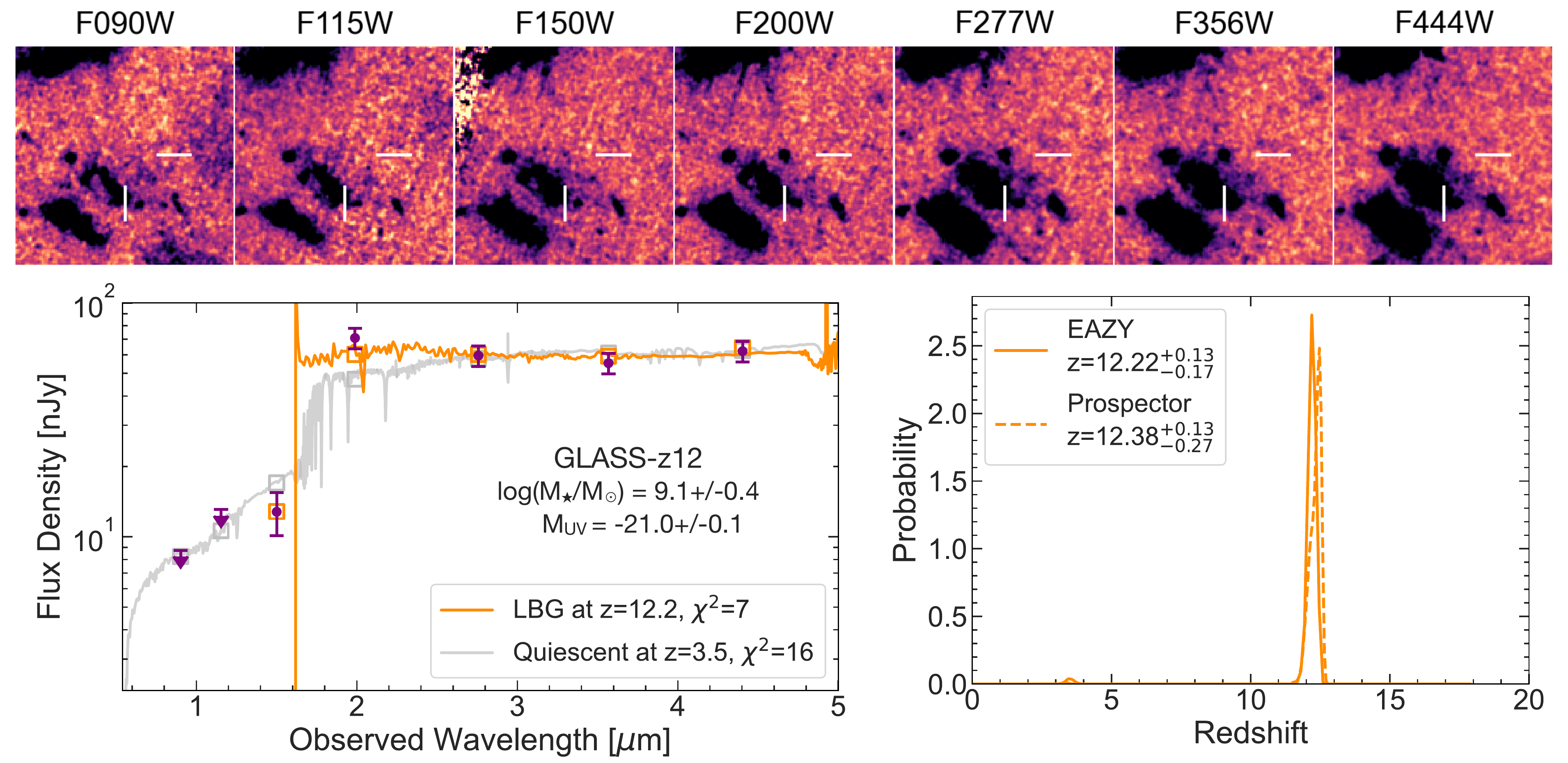}
\caption{Summary of photometry and redshift solution for GL-z12. \textbf{Top:} 4.5''$\times$4.5'' images spanning $\approx0.9-4.5\mu$m centered on the $z\approx12$ candidate highlighted with white crosshairs. The source is well-detected ($>20\sigma$) in F200W and all redder bands, and abruptly drops out in the bluer filters. \textbf{Bottom left:} Photometry for the source is shown in purple, with upper limits for non-detections plotted at the 1-$\sigma$ level. The best-fit spectral energy distribution (SED) template from \texttt{EAZY} is shown in dark orange -- a Lyman-break galaxy (LBG) at $z=12.2$. The best-fit SED from \texttt{EAZY} constrained to lie at $z<6$ is plotted in silver, which corresponds to a quiescent galaxy at $z\approx3.5$ whose Balmer break produces a drop-off across F200W and F150W. However, such a quiescent galaxy is predicted to be detected ($>5\sigma$) in bluer bands, and is at odds with the dramatic $>1.8$ mag break observed. \textbf{Bottom right:} Probability distributions for the source redshift derived using \texttt{EAZY} (solid orange) and \texttt{Prospector} (dashed orange). We adopt a flat prior across the redshift range depicted ($z=0-20$). The derived distributions are in excellent agreement and suggest a redshift of $z\approx12$, with negligible (\texttt{EAZY}) or no (\texttt{Prospector}) support for solutions at $z<10$}.
\label{fig:summaryGLz13}
\end{figure*}

\begin{figure*}
\centering
\includegraphics[width=0.95\linewidth]{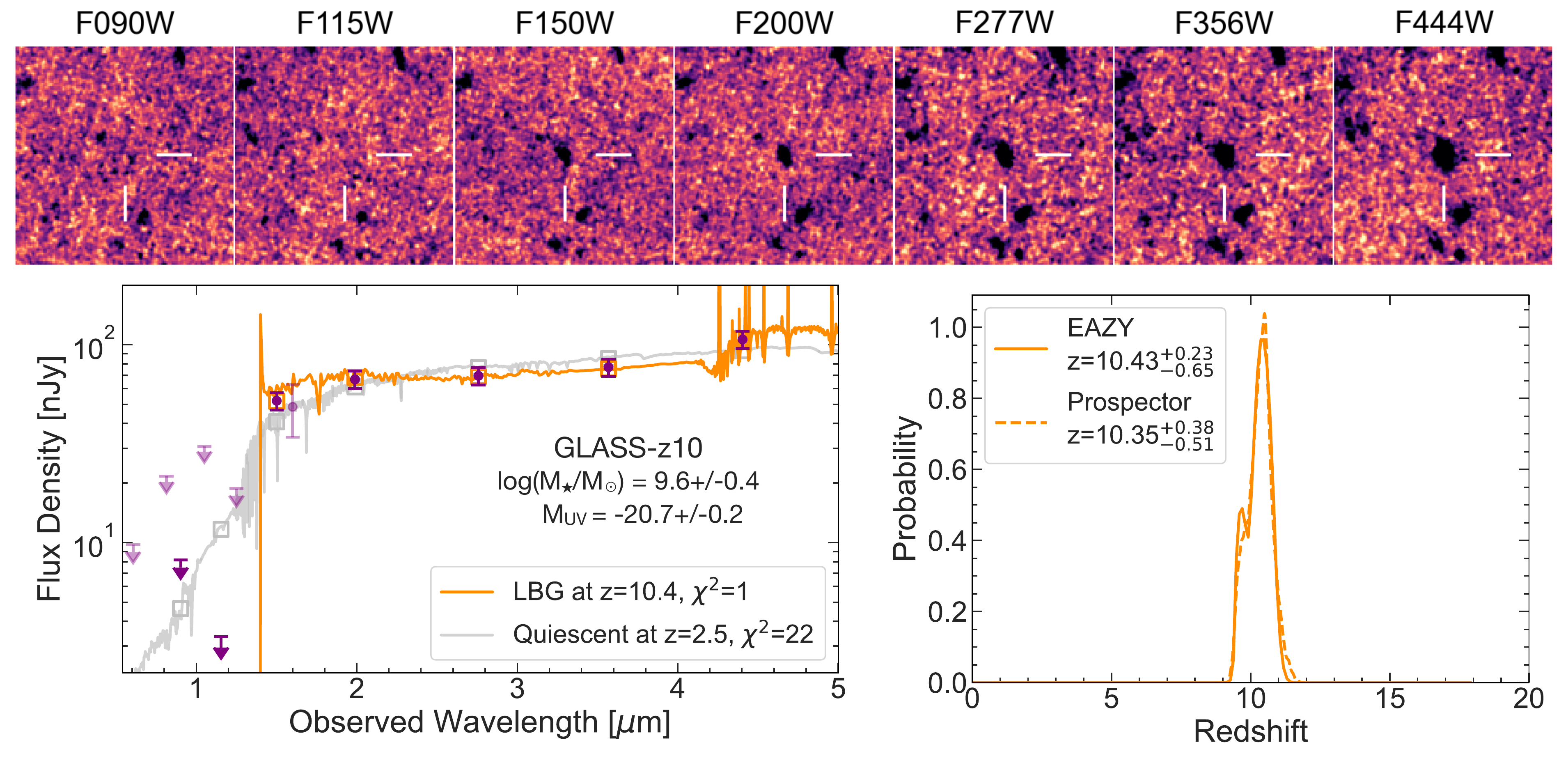}
\caption{Summary of photometry and redshift solution for GL-z10, similar to Figure 1. \textbf{Top:} GL-z10 is well-detected in all but the two bluest bands. \textbf{Bottom left:} The best-fit low-$z$ solution (quiescent galaxy at $z\approx2.5$) is disfavored by the F115W image, where a $>5\sigma$ detection is expected. In addition to the JWST data (dark purple), we measure HST photometry (light purple) for this source from data acquired by the BUFFALO program \citep[][]{Steinhardt20}. The HST data are fully consistent with the JWST data as well as the best-fit SED. \textbf{Bottom right:} The \texttt{EAZY} and \texttt{Prospector} posteriors agree on a $z\approx10$ galaxy.}
\label{fig:summaryGLz11}
\end{figure*}

\begin{figure*}
\centering
\includegraphics[width=0.8\linewidth]{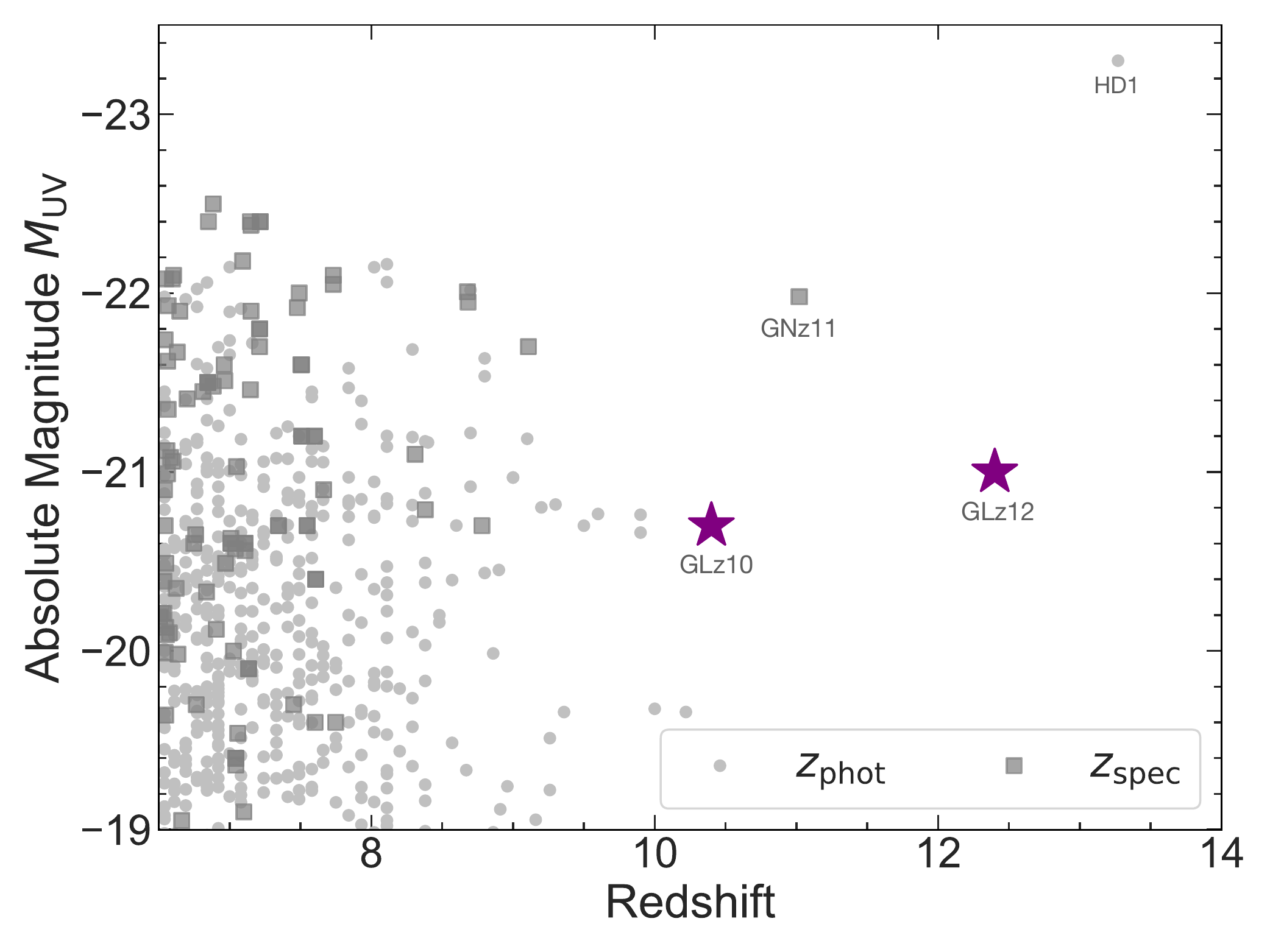}
\caption{Absolute UV magnitude vs. Redshift for a representative sample of known galaxies in the first billion years of the Universe. Galaxies with photometric redshifts, sourced from \citet[][]{Bouwens22}, are shown as points, and those with spectroscopic redshifts compiled from the literature as squares. The candidates presented in this work are depicted as purple stars, and populate a hitherto unoccupied region of parameter space. The brightness of these sources present a unique opportunity to efficiently extend the spectroscopic frontier to the first few hundred Myrs after the Big Bang.}
\label{fig:Muv}
\end{figure*}

\section{Data}
\label{sec:data}

\subsection{Early Extragalactic JWST Observations}

Our analysis is based on some of the first JWST/NIRCam datasets that have been observed and released over extragalactic fields. In particular, we analyze the two Early Release Science programs GLASS and CEERS. 

The first is the NIRCam parallel field of the ERS program GLASS (PID: 1324, \citealt{Treu22}). 
This program obtained a single NIRCam pointing in seven wide filters F090W, F115W, F150W, F200W, F277W, F356W, and F444W, observed for 3.3, 3.3, 1.7, 1.5, 1.5, 1.7, and 6.6 hrs, respectively. 



We also analyzed the first four NIRCam pointings from the ERS program CEERS (PID: 1345, Finkelstein et al., in prep.) that have been observed to date. The CEERS images include F115W+F277W, F115W+F356W, F150W+F410M, and F200W+F444W short- (SW) and long-wavelength (LW) exposure sets, for a typical integration time of 0.8 hr per filter, except for F115W that obtained double this integration time. 

The combined area of these five NIRCam fields used in our analysis amounts to 49 arcmin$^2$ ($\approx10$ arcmin$^2$ in GLASS, and $\approx40$ arcmin$^2$ in CEERS) reaching an unprecedented 5$\sigma$ depth that ranges between 28.6 and 29.6 AB mag at 4 $\micron$ as measured in 0\farcs32 diameter apertures (see Table \ref{table:depth}).

\subsection{Data Reduction}

The NIRCam images were reduced with the standard JWST pipeline up to stage 2 using the reference files from jwst\_0942.pmap, as well as our own sky flats. Additional, chip-dependent zeropoint offsets were applied based on our own reductions of the flux standard star J1743045 as well as observations of the Large Magellanic Cloud. For some filters, the individual corrections in different modules of the NIRCam detector can reach up to 30\% \footnote{for more information see: \url{https://github.com/gbrammer/grizli/pull/107}}.
The resulting images were then processed with the \texttt{grizli}\footnote{\url{https://github.com/gbrammer/grizli/}} pipeline for proper alignment onto a common WCS that was matched to the Gaia DR3 catalogs. 
To do this, \texttt{grizli} re-computes the traditional SIP distortion headers that were common for HST data. This allows us to use \texttt{drizzlepac} as with HST images to combine the individual frames and produce distortion-corrected mosaics.
Additionally, \texttt{grizli} mitigates the 1/f noise and masks `snowballs' that are most prominent in short-wavelength filters.

The pipeline was used to drizzle images at 20mas / 40mas pixels for the SW and LW data, respectively. In the following, our analysis is based on pixel-matched 40mas images, however. 
For more information on the \texttt{grizli} processing see Brammer et al., (in prep).

\subsection{Multi-Wavelength Catalogs}

After processing the new NIRCam images, we produced photometric catalogs in all fields using the \texttt{SExtractor} software \citep[][]{Bertin96}. Sources were detected in dual mode with two different detection images: F200W or a weighted combination of all the LW filters. Fluxes were measured in small circular apertures of 0\farcs32 diameter and were corrected to total using the AUTO flux measurement from the detection image. An additional correction of typically a few percent only was applied for remaining flux outside of this aperture based on the predicted encircled energy for the JWST point-spread functions. Flux uncertainties were estimated based on sigma-clipped histograms of circular apertures placed throughout the images in random sky positions. These were then used to rescale the drizzled rms maps. Thus, our uncertainties are as close to the data as possible. The 5$\sigma$ depths derived in this way are listed in Table \ref{table:depth}.

\subsection{Quality Control}

Given that JWST is a completely new facility for which calibration is still ongoing, it is important to test the resulting images for any issues. Indeed, the NIRCam data revealed several features that are not accounted for in the standard pipeline. In particular, the SW data suffer from significant scattered light, if there are bright stars in the vicinity. This is particularly pronounced in the GLASS parallel field where a 10th mag star just outside of the field seems to cause artificial images across the field. This is particularly pronounced in the F090W filter in the B4 detector. However, also other detectors and filters seem to be affected, albeit to a lower extent. While this issue could be overcome with improved processing, or with additional observations at a different roll angle, this does not significantly affect the current analysis. Since we  are searching for very high redshift galaxies that disappear at shorter wavelengths, these data issues only introduce some level of incompleteness. Most importantly, the areas of the two candidate sources presented later in this paper are not affected.

Where available, we compared our JWST photometry in the shorter wavelength filters with existing HST data to check both for issues with residual distortion or with magnitude zeropoint offsets. In particular, for the GLASS parallel field, we used HST images made available by \citet{Kokorev22} from the BUFFALO survey \citep{Steinhardt20}. However, only a small portion of the field is covered by both HST and JWST. For the CEERS data, we make use of re-reductions from the imaging taken by the CANDELS survey \citep[e.g.,][]{Koekemoer12}.
No significant offsets or issues were detected. Zeropoint corrections remained small ($<15\%$; see next section).

\begin{deluxetable}{lrrrrrrrrrrr}
\label{table:depth}
\tabletypesize{\footnotesize}
\tablecaption{5$\sigma$ Depth of JWST Data in this Analysis}
\tablehead{
\colhead{Band} & \colhead{GLASS-ERS} &  \colhead{CEERS-ERS}
}
\startdata
\vspace{-0.2cm}  \\
$F$090$W$ & 29.0  & \nodata \\
$F$115$W$ & 29.0  & 28.8\\
$F$150$W$ & 28.9  & 28.6\\
$F$200$W$ & 29.1  & 28.8\\
$F$277$W$ & 29.4  & 28.9\\
$F$356$W$ & 29.4  & 29.1\\
$F$410$M$ & \nodata  & 28.4\\
$F$444$W$ & 29.6  & 28.7 \\
\enddata
\tablecomments{Measured in 0\farcs32 diameter circular apertures.}
\end{deluxetable}

\begin{deluxetable}{lrrrrrrrrrrr}
\label{table:photometry}
\tabletypesize{\footnotesize}
\tablecaption{Photometry in units of nJy}
\tablehead{
\colhead{Band} & \colhead{GL-z10} & \colhead{GL-z12} 
}
\startdata
\vspace{-0.2cm}  \\
$F$090$W$ & 4$\pm$4 & 5$\pm$4\\
$F$115$W$ & 1$\pm$3 & 8$\pm$5\\
$F$150$W$ & 52$\pm$3 & 13$\pm$3\\
$F$200$W$ & 67$\pm$2 & 71$\pm$2\\
$F$277$W$ & 68$\pm$2 & 59$\pm$2\\
$F$356$W$ & 71$\pm$2 & 51$\pm$2\\
$F$444$W$ & 98$\pm$2 & 57$\pm$2\\
\enddata
\tablecomments{We set an error floor of $10\%$ on our measured fluxes for \texttt{EAZY} and \texttt{Prospector} fits to account for systematic uncertainty not reflected in the errors stated above.}
\end{deluxetable}

\begin{deluxetable}{lrrrrrrrrrrr}
\label{table:properties}
\tabletypesize{\footnotesize}
\tablecaption{Summary of properties.}
\tablehead{
\colhead{} & \colhead{GL-z10} & \colhead{GL-z12} 
}
\startdata
\vspace{-0.2cm}  \\
R.A. & +0:14:02.86 & +0:13:59.76\\
Dec. & $-$30:22:18.7 & $-$30:19:29.1\\
Redshift $z_{\rm{Prospector}}$ & $10.4^{+0.4}_{-0.5}$  & $12.4^{+0.1}_{-0.3}$\\
Redshift $z_{\rm{EAZY}}$ & $10.4^{+0.2}_{-0.7}$ & $12.2^{+0.1}_{-0.2}$\\
Stellar Mass $\log$($M_{\rm{\star}}/M_{\rm{\odot}}$) & $9.6^{+0.2}_{-0.4}$ & $9.1^{+0.3}_{-0.4}$\\
UV Luminosity ($M_{\rm{UV}}$) & $-20.7^{+0.2}_{-0.2}$  & $-21.0^{+0.1}_{-0.1}$\\
UV Slope ($\beta$; $f_{\rm{\lambda}}\propto \lambda^{\beta}$) & $-1.9^{+0.2}_{-0.1}$ & $-2.3^{+0.2}_{-0.2}$\\
Dust Attenuation ($A_{\rm{5500\AA}}$) & $0.3^{+0.4}_{-0.2}$ & $0.1^{+0.2}_{-0.1}$\\
Dust Attenuation ($A_{\rm{1500\AA}}$) & $0.8^{+0.7}_{-0.5}$ & $0.3^{+0.4}_{-0.2}$\\
Age ($t_{\rm{50}}$/Myr) & $163^{+20}_{-133}$ & $111^{+26}_{-83}$ \\
SFR$_{\rm{50\ Myr}}$ ($M_{\rm{\odot}}$/yr) &  $10^{+17}_{-5}$ & $6^{+5}_{-2}$\\
$r_{\mathrm{eff}}$ [kpc] & 0.7  & 0.5 \\ 
Sersic Index $n$ & 0.8  & 1.0 \\ 
\enddata
\tablecomments{SED fitting assumes a continuity prior on the star-formation history and a \citet[][]{Chabrier03} IMF.}
\end{deluxetable}

\section{Sample Selection \& Methods}
\label{sec:Sample}


Photometric redshifts form the basis of our search for bright $z>10$ galaxies. We fit redshifts using \texttt{EAZY} \citep{Brammer08} adopting the ``tweak\_fsps\_QSF\_12\_v3" template set derived from \texttt{FSPS} \citep{FSPS1,FSPS2,FSPS3}. The allowed range is $0.1<z<20$ adopting a flat luminosity prior, after applying modest ($<15\%$) zero-point corrections that are derived iteratively. Candidates of interest are selected to have best-fit redshifts $z>10$ along with $>84\%$ of their derived redshift probability distribution function, $p(z)$, lying at $z>10$. 

We further require $>10\sigma$ detections in both F356W and F444W. These bands sample the rest-frame UV at $z>10$ and are critical in establishing the flux levels with respect to which we seek strong Lyman breaks.


We inspect images of every candidate source for data quality issues (e.g., contamination from neighbors, diffraction spikes, location on the edge of the detector). In tandem, we examine plausible low-$z$ solutions for the candidates by running \texttt{EAZY} constrained to $z<6$ (primarily dusty, quiescent galaxies with Balmer breaks masquerading as Lyman breaks). We inspect low-$z$ solutions with the understanding that the errors on fluxes in the dropout band may be underestimated in some cases (in e.g., image areas with residual striping), which can have an important impact on the $p(z)$. For instance, we find one bright source in CEERS with a confident $z_{\rm{EAZY}}\approx17$ and $z_{\rm{Prospector}}\approx17$ whose $z>10$ solution assumes a secure non-detection in F150W -- unfortunately, the source falls in a low-SNR region of the F150W image and it is difficult to judge the reality of the non-detection.

We find 5 plausible $z>10$ candidates that survive all our conservative checks -- 3 in CEERS, and 2 in GLASS. Of these, the two identified in GLASS -- GL-z10 ($z_{\rm{EAZY}}=10.4^{+0.2}_{-0.7}$) and GL-z12 ($z_{\rm{EAZY}}=12.4^{+0.2}_{-0.2}$) --  stand out as being particularly luminous and secure. No other objects when fit with the \texttt{Prospector} SED-fitting code (see following Section) have a $p(z)$ with all modes contained at $z>10$. Further, the GLASS candidates are among the most luminous found -- GL-z10, in particular, is by far the brightest of all sources (by $>2\times$ in F444W). For the rest of this work, we focus on these two particularly luminous candidates and we defer the rest of the sources to future papers that present an analysis of the full $z>10$ galaxy population in these fields.

\section{Results}
\label{sec:results}
\subsection{Two Luminous $z>10$ Galaxy Candidates}
\label{sec:candidates}

We confirm the photometric redshifts for the two GLASS candidates and derive stellar population properties using the \texttt{Prospector} SED fitting code \citep[][]{Leja17,Leja19,Johnson21}. The SED parameter space explored by \texttt{Prospector} is more expansive than \texttt{EAZY}'s linear template combinations, and therefore it acts as an important check on our derived redshifts. We use FSPS \citep{FSPS1,FSPS2,FSPS3} with the MIST stellar models \citep[][]{Choi17}. We adopt the 19-parameter physical model and parameter choices described in \citet{Tacchella22} that fits for the redshift, stellar and gas-phase metallicities, stellar mass, star-formation history, dust properties, AGN emission, and scaling of the IGM attenuation curve. We make slight modifications to their setup -- in particular we explore a broader redshift range of $z=0.1-20$ and keep two bins fixed at lookback times of $0-5$ Myr and $5-10$ Myrs in the star-formation history following \citet{Whitler22} to capture recent bursts that may be powering extreme nebular emission expected to occur generically at the redshifts of interest \citep[e.g.,][]{Labbe13,debarros19,Endsley19}. We adopt a ``continuity" prior on the star-formation history, which limits the amount of variance across consecutive time-bins resulting in smooth histories \citep[][]{Leja19,Tacchella22}. For further details we direct readers to Table 1 and \S3.4 of \citet{Tacchella22}.

The redshift fits from \texttt{Prospector} are in excellent agreement with \texttt{EAZY} -- we find $z_{\rm{Prospector}}= 10.4^{+0.4}_{-0.5}$ for GL-z10 and $z_{\rm{Prospector}}= 12.4^{+0.1}_{-0.3}$ for GL-z12. The photometry and redshift inference for these sources are summarized in Figures \ref{fig:summaryGLz13} and \ref{fig:summaryGLz11}, with fluxes listed in Table \ref{table:photometry}. We confirm that no significant data quality issues affects the $z>10$ candidacy of the sources in the imaging. We derive $p(z>10)\approx100\%$ for GL-z12 and $p(z>9.4)\approx100\%$ for GL-z10, with their dramatic $\gtrsim1.8$ mag breaks explained by total absorption of photons bluewards of Lyman-$\alpha$ by neutral Hydrogen in the intergalactic medium. 

Both galaxies are detected at very high significance in all filters longward of their break, by virtue of our selection. While they appear very luminous in the JWST data, these sources have UV absolute magnitudes ($M_\mathrm{UV}\approx-21$) that correspond to $L_\mathrm{UV}^*$ at $z\sim8-10$ \citep[see, e.g.,][]{Bouwens21}. This also makes them 1\,mag fainter than GN-z11 and even 2.5 mag fainter than the possible $z\sim13$ galaxy candidate HD1 \citep{Harikane22}. Hence, these sources are not really extreme outliers (see also Fig. \ref{fig:Muv}). Nevertheless, it is interesting that the first few images with JWST already reveal two such bright sources. We will discuss their implications on the UV LF in a later section.

\subsection{Possible Lower Redshift Contamination}
The non-detections of both sources in deep, shorter wavelength images essentially rules out a lower redshift solution. Nevertheless, it is interesting to explore the nature of possible contaminants. We thus rerun our photometric redshift codes and force them to find lower redshift fits. 
The best $z<6$ solutions in our low-$z$ \texttt{EAZY} runs for these sources are $\approx10^{8-9} M_{\rm{\odot}}$ quiescent galaxies at $z\approx3.4$ ($z\approx2.5$) with Balmer breaks straddling the dropout filter (silver SEDs in Figs. \ref{fig:summaryGLz13}, \ref{fig:summaryGLz11}). However note that Balmer breaks, even in the most pathological cases (e.g., 4000\AA\ falls just redward of the dropout filter in a super-solar metallicity galaxy as old as the age of the Universe at $z\approx2-3.5$), can only produce drops of $\lesssim1.5$ mag (assuming no attenuation). The best-fit low-$z$ solutions have $A_{\rm{V}}\approx0.1$ -- stronger attenuation that deepens the break is disfavored by the blue continuum slope at wavelengths longer than the break. In other words, the best-fit low-$z$ solutions predict $>5\sigma$ detections in bands where we find no flux, and continuum slopes redder than we observe.

In order to allow for possible systematic effects in the new JWST data, we perform further testing. We re-fit redshifts to multiple versions of photometry for these sources -- e.g., by adding PSF corrections using \texttt{WebbPSF}, by increasing the error floor on the photometry, by extracting photometry using different apertures and detection bands. The only test that produces viable low-$z$ solutions is when we set a 10 nJy error-floor on all photometry -- this is roughly the level in the SW filters at which the strongest Balmer breaks at $z\approx2-3$ can no longer be ruled out (see open silver squares in bottom-left panels of Figures  \ref{fig:summaryGLz13} and \ref{fig:summaryGLz11}). This test is a vivid demonstration of why the sensitivity of JWST/NIRCam is required to identify objects like GL-z10 and GL-z12 with confidence. 

\subsection{Physical Properties -- Billion $M_{\rm{\odot}}$ Galaxies within $\approx400$ Myrs of the Big Bang}
\label{sec:physical}

While the discovery of GN-z11 has already demonstrated that the formation of billion solar mass galaxies was well underway at $\sim$400 Myr after the Big Bang, the discovery of these two new sources allows us to derive further constraints on the physical properties of galaxies at this very early epoch of the Universe. The \texttt{Prospector} results are summarized in Table \ref{table:properties}. 
In order to efficiently sample the redshift range of interest, we assume a tighter redshift prior (a Gaussian centred on the \texttt{EAZY} $p(z)$ with width set to the 84$^{\rm{th}}$ - 16$^{\rm{th}}$ percentile) than in our previous runs when fitting for the redshift. 

The stellar mass for both objects is constrained to be $\approx10^{9}M_{\rm{\odot}}$, comparable to GNz11 \citep[][]{Oesch16,Johnson21,Tacchella22}. We have verified the stellar mass is stable to changes in the star-formation history prior by also testing the ``bursty" prior from \citet[][]{Tacchella22} which allows more rapid fluctuations in the SFH from time-bin to time-bin than the fiducial model. The star-formation rates averaged over the last 50 Myrs (SFR$_{\rm{50}}$) are typical for galaxies of comparable mass at $z\approx7-10$ \citep[e.g.,][]{Stefanon21}. The SEDs are consistent with negligible dust attenuation and have blue UV slopes, $\beta\lesssim-2$. We note that all these derived properties from the SED fits are collectively consistent with a $z>10$ interpretation for these galaxies.

\begin{figure*}
\centering
\includegraphics[width=0.95\linewidth]{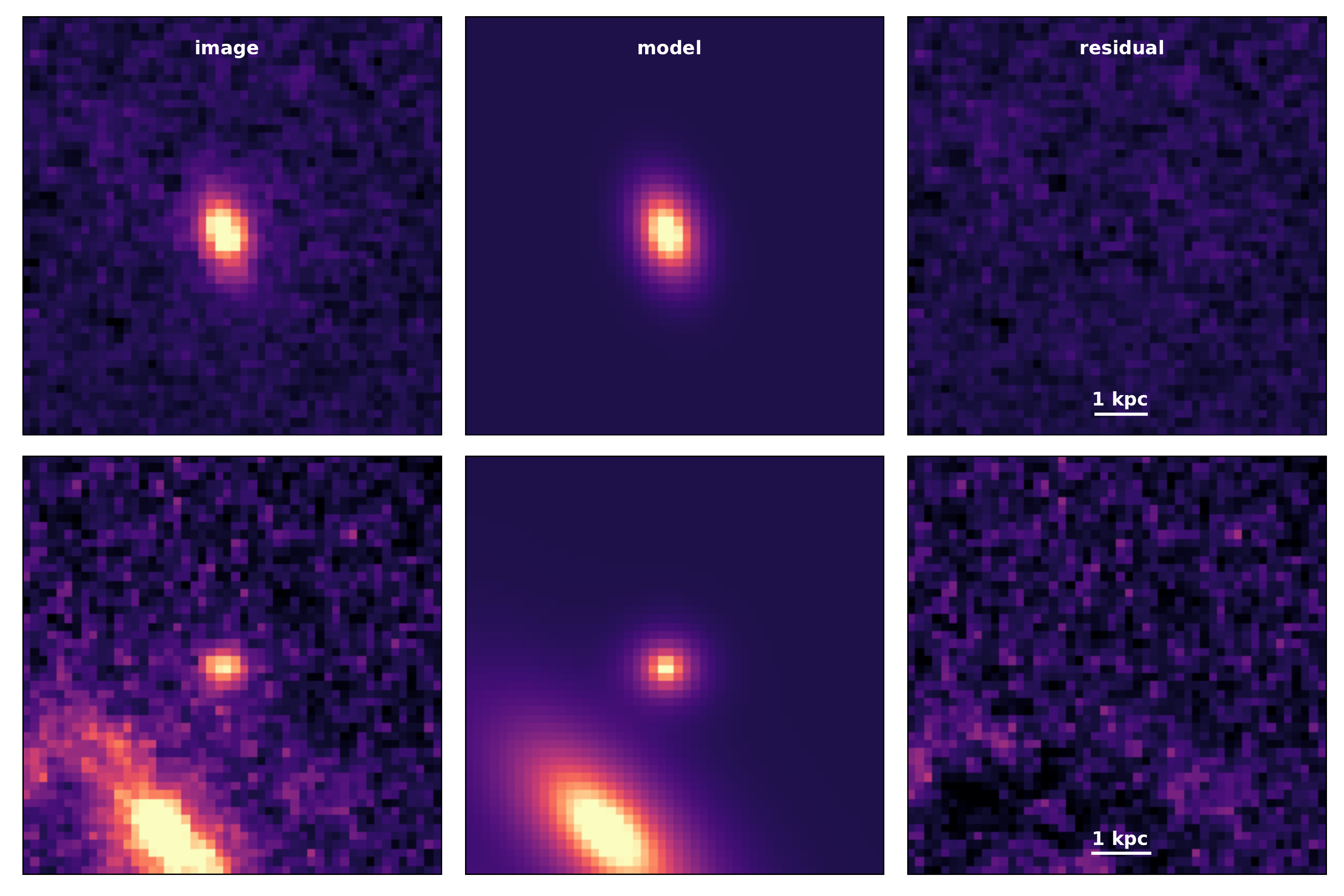}
\caption{Results of the \texttt{GALFIT} morphology analysis for our two sources (GL-z10 top and GL-z12 bottom). The different columns from left to right correspond to the original data (in the F444W filter), the model, and the residual. The sizes and Sersic profiles of both sources are well constrained. GL-z10 shows some clear extension, consistent with a disk galaxy of 0.7 kpc at $z\sim11$. GL-z12 appears quite compact with an estimated size of 0.5 kpc.}
\label{fig:sizefits}
\end{figure*}

\subsection{Galaxy-galaxy Lensing}
\label{sec:lensing}

Galaxy-galaxy lensing may be particularly important at the redshift frontier where flux-limited surveys may be preferentially sampling magnified sources \citep[e.g.,][]{Wyithe11}. Here we make a simple estimate of how lensed our sources are by assuming their neighbors are singular isothermal spheres \citep[e.g.,][]{FortMellier94,Schneider06,Treu10} following e.g., \citet[][]{McLure06,Oesch14,Matthee17lensing}. For this estimate, galaxy redshifts and stellar masses are based on our \texttt{EAZY} fits. Velocity dispersions that trace the underlying dark matter halos are inferred from the stellar mass by extrapolating the empirical scaling relation in \citet[][]{Zahid16} that is fit to $z<0.7$ quiescent galaxies that span $M_{\rm{\star}}\approx10^{9}-10^{12} M_{\rm{\odot}}$. We choose a local relation to cover the low masses relevant to the most massive neighbors at $<10''$ that are likely to produce significant magnification, while noting that the redshift evolution of such relations at least for $M_{\rm{\star}}\gtrsim10^{11} M_{\rm{\odot}}$ is expected to be gradual -- e.g., $\approx20\%$ higher dispersion at fixed stellar mass at $z\approx2$, \citep[e.g.,][]{Mason15lensing}.

For both GL-z10 and GL-z12 we find negligible lensing ($\mu<1.1$) from all foreground sources within 10\farcs. GL-z12 has two relatively massive $M_{\rm{\star}}\approx10^{9} M_{\rm{\odot}}$ neighbors apparent in the bottom-left quadrant of the stamps in Fig. \ref{fig:summaryGLz13}. Even for these two nearby neighbors ($z\approx2$ at a separation of 0\farcs8, and $z\approx3$ at 2\farcs0) the lensing is modest. We further note that GL-z13 has a compact morphology (\S\ref{sec:sizefits}) that does not show elongation along any particular direction that would hint at strong magnification. Based on these considerations we conclude that the observed luminosities of GL-z10 and GL-z12 are likely to be their intrinsic luminosities.

\subsection{The Sizes of Luminous $z\approx10-12$ Galaxies}
\label{sec:sizefits}

We fit the sizes of both candidates in the F444W imaging ($\lambda_{\mathrm{rest}} \sim 3500 \mathrm{\AA}$) using \texttt{GALFIT} \citep{Peng10Galfit}. We create 100-pixel cutouts around each galaxy, then use \texttt{photutils} and \texttt{astropy} to create a segmentation map to identify nearby galaxies. We simultaneously fit any sources with magnitudes (estimated from the segmentation map) up to 2.5 magnitudes fainter than the target galaxy that have centers within 3\arcsec of the galaxy; we mask fainter or more distant galaxies. In our fits, we constrain the center of the target galaxy to be within 10 pixels (0\farcs4) of the input value, the Sersic index $n$ to be between 0.01 and 8, the magnitude to be between 0 and 45, and the half-light radius $r_e$ to be between 0.3 and 200 pixels (0\farcs012 - 8\farcs0). We calculate and subtract off a scalar sky background correction from each cutout, estimated from the masked, sigma-clipped cutout, then fix the sky background component in \texttt{GALFIT} to zero. We use a theoretical PSF model generated from WebbPSF at our 0\farcs04 pixel scale; we oversample the PSF by a factor of 9 in order to minimize artifacts as we rotate the PSF to the GLASS observation angle calculated from the APT file, then convolve with a 9x9 pixel square kernel and downsample to the mosaic resolution. 

We find reliable Sersic fits for both galaxies, with half-light radii of 0.5 and 0.7 kpc, respectively, and disk-like profiles ($n=1$ and $n=0.8$, respectively). The models are shown in Fig. \ref{fig:sizefits}, and the size and Sersic profile estimates are listed in Table \ref{table:properties}.

The resulting sizes of 0.5 and 0.7 kpc are typical for luminous $L^*$ galaxies at $z\sim6-9$, where measurements have been possible to date \citep[e.g.,][]{Holwerda15,Shibuya15,Bowler17,Kawamata18,Yang22}.
They are also consistent with expectations from simulations for $z>9$ galaxies \citep[e.g.,][]{Roper22,Marshall22}.
However, at $z\sim7$, the most luminous sources often break up in multiple clumps \citep{Bowler17}. This is not the case for these two sources, at least down to the resolution limit of order 500 pc for the F444W bandpass. Interestingly, GL-z10 even shows tantalizing evidence for being an ordered disk galaxy at $z\sim10$, based on the exponential light profile and elliptical morphology. If we interpret GL-z10's projected axis ratio of 0.65 using a sample of randomly oriented axisymmetric oblate rotators (following e.g., \citealt{Holden12}, \citealt{Chang13}, \citealt{vanderWel14b}) and adopt $c/a\leq0.4$ as a threshold for disks, we find that the observed axis ratio implies P(disk) $\sim0.5$. Our analysis shows the unparalleled  power of JWST to provide accurate profile measurements of early Universe galaxies.

\section{Discussion}
\label{sec:discussion}

\subsection{Caveats}
\label{sec:caveats}
The key caveat, as well as the key animating spirit of this work, is that these data are among the first deep extragalactic fields collected by a new Great Observatory. Systematic uncertainties (e.g., zero-point corrections, treatment of artefacts) can still be significant. We have tested for zero-point offsets by comparing HST and JWST photometry for brighter sources where possible and have not found any major issues at the $\lesssim 10\%$ level. Nevertheless, we have attempted to account for remaining uncertainties with conservative choices -- e.g., a $10\%$ error floor on all fluxes and focusing on bright galaxies whose $>2$ mag breaks are robust to even major uncertainties.

A next caveat applies to the SED models underpinning the stellar population parameters and photo-$z$ fitting. Important details about the nebular emission and nature of massive stars at low metallicities, which dominate the light in these few hundred Myr old star-forming systems, remain unconstrained \citep[e.g.,][]{Stanway20}. These uncertainties directly translate to the parameters we recover from SED fitting and the sources for which we are able to fit high quality redshifts. Fortunately, for our redshift range of interest, from the perspective of continuum fitting for the stellar population analyses, extreme nebular emission from strong rest-frame optical lines is shifted out of all NIRCam bands and the most important feature for the photo-$z$s in the bright galaxies we study is the Lyman break. Further, there exists a range of plausible, but hitherto unconstrained physical ingredients that are unaccounted for in our models (e.g., primordial AGN, top-heavy IMFs, super-luminous Pop III stars;  \citealt{Windhorst18,Pacucci22,Steinhardt22}). Some of these ingredients may potentially produce large UV luminosities in the absence of substantial stellar mass. Spectroscopic follow-up is therefore essential to confirm the nature of these sources.


\begin{figure*}
\centering
\includegraphics[width=0.8\linewidth]{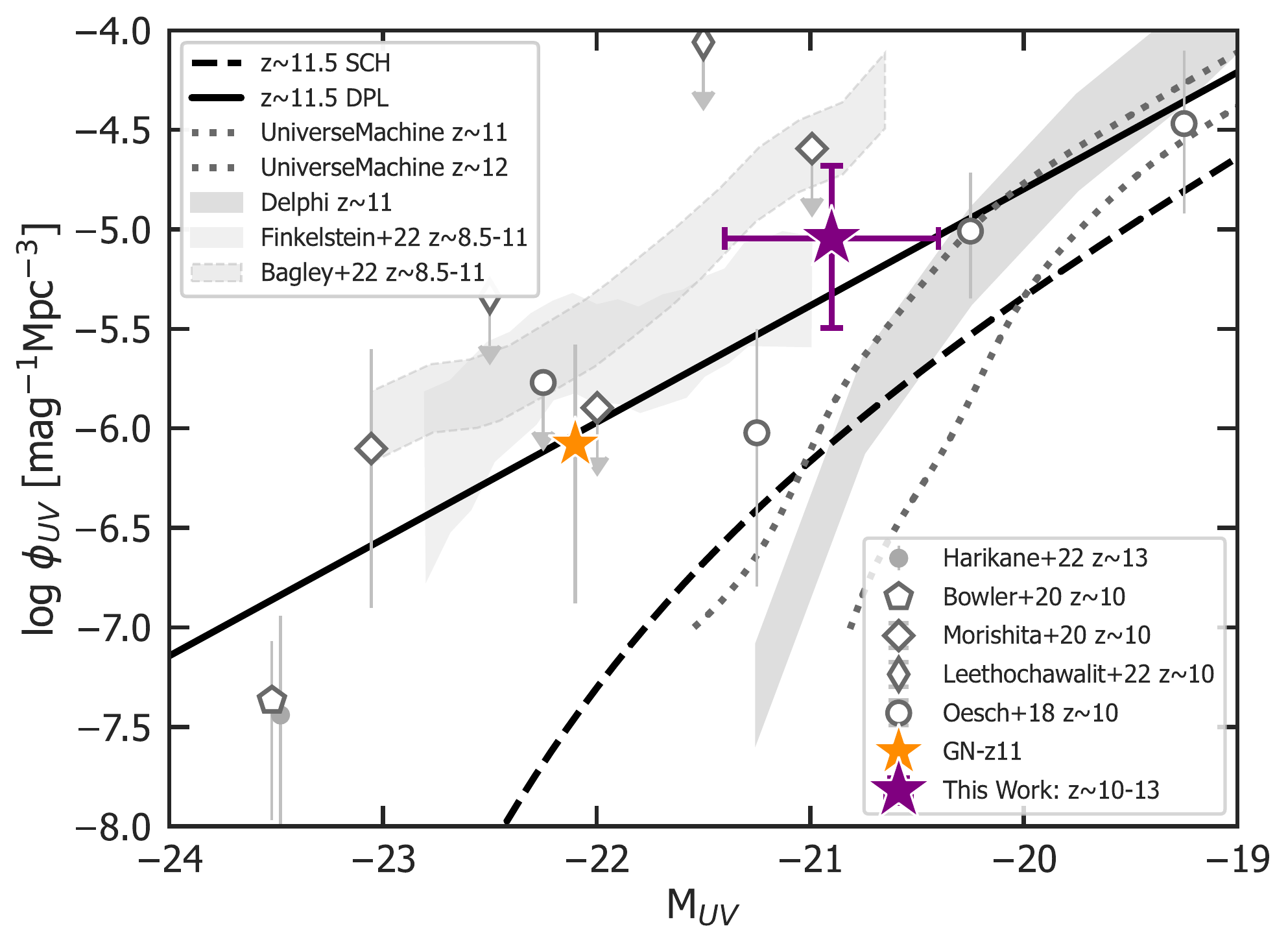}
\caption{Constraints on the bright end of the UV LF at $z\sim10-13$. The current JWST data allow us to derive a first estimate of the number density of galaxies with $M_\mathrm{UV}\sim-21$ at these redshifts (purple star). While this estimate lies a factor $\sim10\times$ above the extrapolation of Schechter function constraints to $z=11.5$ from \citet[][dashed black line]{Bouwens21}, they are in very good agreement with extrapolated double-power law LFs from \citet[][black solid line]{Bowler20}. In fact, GN-z11 (orange star) is also consistent with the double-power law LF. Other LF estimates and upper limits at $z\sim10$ are shown as open symbols (see legend for references). Simulated predictions are shown from the UniverseMachine models at $z\sim11$ and $z\sim12$ (dotted lines) and from the Delphi model at $z\sim11$ (darker gray shaded region).}
\label{fig:UVLFtest}
\end{figure*}

\subsection{Implication: The Number Density of Luminous Galaxies in the Early Universe}
\label{sec:numberdensity}
These two galaxies enable a first estimate of the number densities of relatively luminous sources at $z=10-13$. Given the large uncertainties in redshift, we conservatively estimate a selection volume across this full redshift range, i.e., $\Delta z=3$. Given that the depth of all the fields we studied is far deeper than the bright magnitude of these two sources, these galaxies would have been identified over the full area, without foreground galaxies.  This amounts to a volume of $2.2\times10^5$ Mpc$^3$. The detection of two galaxies with $M_\mathrm{UV}=-21$ thus results in an estimated UV LF point of $\log\phi[\mathrm{Mpc}^{-3}\mathrm{mag}^{-1}]=-5.05^{+0.37}_{-0.45}$. In \S\ref{sec:lensing} we conclude that this UV LF point likely reflects the intrinsic luminosity of the sources, and is unaffected by lensing.

In Figure \ref{fig:UVLFtest}, we compare this estimate with previous UV LF determinations and with extrapolations from lower redshifts. In particular, we show that an extrapolation of the Schechter function trends estimated at $z=3-10$ results in an LF at $M_\mathrm{UV}=-21$ that is a factor $>$10$\times$ lower than our estimate. 
Interestingly, however, when extrapolating the trends in the double-power law LFs from \citet{Bowler20} to $z\sim11.5$ (the mean redshift of our sources), we find relatively good agreement. In fact, GN-z11 also lies on this extrapolated LF. 
However, this would indicate very little evolution in the bright galaxy population at $z>8$. Indeed, our estimate is in good agreement with previous $z\sim10$ UV LF determinations and constraints at the bright end from \citet{Oesch18,Bouwens19,Stefanon19,Morishita20,Finkelstein22,Bagley22,Leethochawalit22}. 

Finally, we also briefly compare our estimates with simulated LFs from the UniverseMachine model \citep{Behroozi19} and from the Delphi model \citep[][]{Dayal14,Dayal22}. While our estimate is in rough agreement with these prediction at $z\sim11$, the model LFs evolve very rapidly at these early times, such that the $z\sim12$ LF is already $>30\times$ below our estimate. This is a general trend of model predictions: a relatively rapid evolution of the LF at $z>10$, driven by the underlying evolution of the dark matter halo mass function \citep[see also][]{Oesch18, Tacchella18,Bouwens21}. However, the handful of bright galaxies that have been found at $z\sim10-13$ to date appear to oppose this trend. 

It is still unclear what the physical reason for this might be. Combined with the discovery of GN-z11 \citep[][]{Oesch16}, and taking our SED fits at face value (see \S\ref{sec:caveats} for caveats), evidence is mounting that the star-formation efficiency in the early Universe may be much higher than expected (e.g., \citealt{Tacchella13,Mason15,Tacchella18}) in at least a few sources, thus resulting in the early appearance of UV-luminous galaxies with stellar masses as high as $\approx10^{9}M_{\rm{\odot}}$ already a few hundred Myrs after the Big Bang. The existence of these massive galaxies at such early times raises interesting questions about just how early such galaxies began forming, potentially earlier than current expectations. Wider area datasets will be required to increase the search volume, for more reliable constraints on the number densities of luminous sources. 


\section{Summary \& Outlook}
\label{sec:summary}

This paper presented a search for luminous $z>10$ galaxies across the two JWST Early Release Science programs in extragalactic fields. We find the following -- 

\begin{itemize}

\item We identify two particularly luminous sources in the GLASS ERS program.
These sources, GL-z10 and GL-z12, have continuum magnitudes of $\sim27$ at 2 $\micron$ and display dramatic $>1.8$ mag breaks in their SEDs that are best fit as Lyman breaks occurring at redshifts of $z\approx11$ and $z\approx13$ respectively. [Fig. \ref{fig:summaryGLz13}, Fig. \ref{fig:summaryGLz11}, \S\ref{sec:candidates}]

\item SED modeling of these sources shows they have properties (e.g., $\beta$ slopes, specific star-formation rates) expected of $z>10$ galaxies. These systems are a billion solar mass galaxies, having built up their mass only $<300-400$ Myrs after the Big Bang. [Table \ref{table:properties}, \S\ref{sec:physical}]

\item The brightness of these objects present a unique opportunity for detailed spectroscopic and morphological follow-up at $z>10$. As a demonstration, we model the morphology of both galaxies finding that they are well-described by disk-like profiles with small sizes (half-light radii $\sim0.6$\ kpc). GL-z10, in particular, shows an extended exponential light profile, that may be tantalizing evidence for a disk already in place at $z\approx10$.
[Fig. \ref{fig:sizefits}, \S\ref{sec:sizefits}]

\item These two objects already place novel constraints on galaxy evolution in the cosmic dawn epoch. They indicate that the discovery of GN-z11 was not simply a matter of good fortune, but that there is likely a population of UV luminous sources with very high star-formation efficiencies capable of compiling $>10^{9} M_{\rm{\odot}}$ at $z>10$. [Fig. \ref{fig:Muv}, \S\ref{sec:numberdensity}]

\item The inferred number-density of $M_{\rm{UV}}\approx-21$ sources from our search ($\log\phi[\mathrm{Mpc}^{-3}\mathrm{mag}^{-1}]=-5.05^{+0.37}_{-0.45}$) strongly supports a significant deviation from the Schechter UV luminosity function at the bright end, and is consistent with the double-power law evolution reported at lower redshifts. The physical mechanisms driving this departure are yet to be definitively established. These luminous sources highly conducive to NIRSpec spectroscopy may hold the key. [Fig. \ref{fig:UVLFtest}, \S\ref{sec:numberdensity}]
\end{itemize}

If these candidates are confirmed spectroscopically, and indeed two $z\approx10-12$ candidates lie awaiting discovery in every $\sim$50 arcmin$^{2}$ extragalactic field, it is clear that JWST will prove highly successful in pushing the cosmic frontier all the way to the brink of the Big Bang.

\facilities{\textit{JWST}, \textit{HST}}

\software{
    \package{IPython} \citep{ipython},
    \package{matplotlib} \citep{matplotlib},
    \package{numpy} \citep{numpy},
    \package{scipy} \citep{scipy},
    \package{jupyter} \citep{jupyter},
    \package{Astropy}
    \citep{astropy1, astropy2},
    \package{grizli}
    (v1.5.0; \citealt{grizli,grizli2}),
    \package{MIST} \citep[][]{Choi17},
    \package{Prospector} \citep[][]{Leja17,Leja19,Johnson21},
    \package{FSPS} \citep[][]{FSPS1,FSPS2,FSPS3,FSPS4,python-FSPS},
    \package{EAZY} \citep[][]{Brammer08},
    \package{SExtractor} \citep[][]{Bertin96},
    \package{GALFIT} \citep[][]{Peng02,Peng10Galfit}
    }
    
\acknowledgments{

We thank the referee for insightful comments that strengthened this manuscript. We are grateful to the CEERS and GLASS teams for planning these early release observations.

RPN acknowledges funding from \textit{JWST} programs GO-1933 and GO-2279. Support for this work was provided by NASA through the NASA Hubble Fellowship grant HST-HF2-51515.001-A awarded by the Space Telescope Science Institute, which is operated by the Association of Universities for Research in Astronomy, Incorporated, under NASA contract NAS5-26555. We acknowledge support from: the Swiss National Science Foundation through project grant 200020\_207349 (PAO, AW). The Cosmic Dawn Center (DAWN) is funded by the Danish National Research Foundation under grant No.\ 140. RJB and MS acknowledge support from NWO grant TOP1.16.057.

Cloud-based data processing and file storage for this work is provided by the AWS Cloud Credits for Research program.

This work is based on observations made with the NASA/ESA/CSA James Webb Space Telescope. The data were obtained from the Mikulski Archive for Space Telescopes at the Space Telescope Science Institute, which is operated by the Association of Universities for Research in Astronomy, Inc., under NASA contract NAS 5-03127 for JWST. These observations are associated with programs \# 1324 and \# 1345.
}

\bibliography{MasterBiblio}
\bibliographystyle{apj}

\appendix
\section{Comparison with Initial Reduction and Calibration}
\label{appendix:oldfluxes}

\begin{figure*}
\centering
\includegraphics[width=0.95\linewidth]{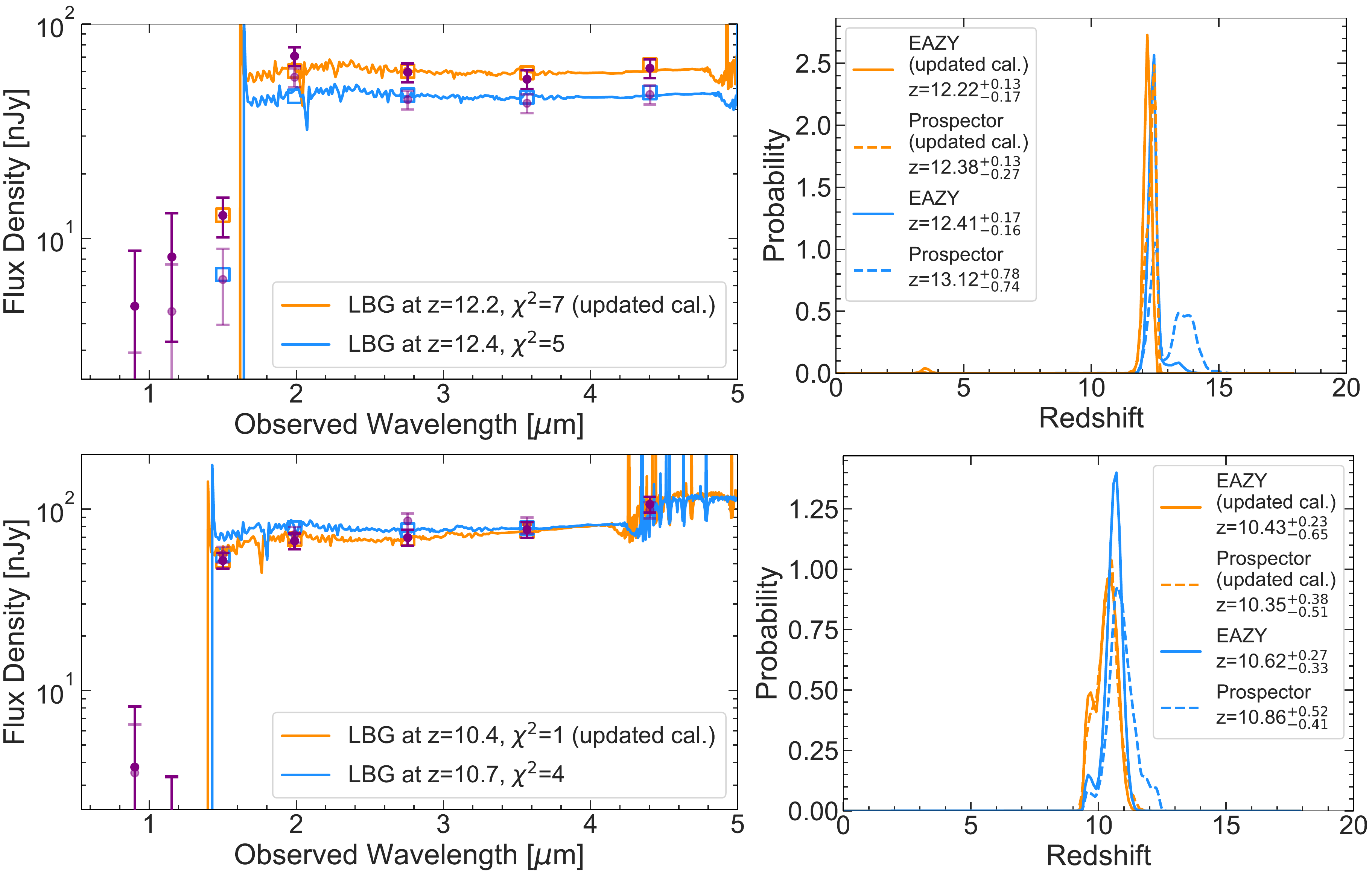}
\caption{Comparison of SEDs (left) and $p(z)$ distributions (right) derived based on initial NIRCam calibrations available at the time of the ERS data release (light blue) with results based on updated calibrations used in this paper (orange). Photometry from the initial calibrations is shown in the left panels in transparent purple, and from the updated calibrations in solid purple. The SEDs are broadly similar, with the updated calibrations implying slightly lower redshifts for both sources. The high redshift nature of these sources remains secure.}
\label{fig:appendix}
\end{figure*}

In Figure \ref{fig:appendix} we compare the results presented in this paper with fluxes, SEDs, and $p(z)$ based on the initial NIRCam calibrations (e.g., zero-points, flats, darks) available with the ERS data release in July 2022. Relative to results based on these calibrations, the $p(z)$ distribution has shifted slightly towards lower redshifts for both galaxies. GL-z10 now shows a stronger hint of rest-optical lines or a Balmer break in its $F$444$W$-$F$356$W$ color, along with a slightly redder $\beta_{\rm{UV}}$ slope ($-1.9$ vs. $-2.1$) which does not require as much damping by the IGM to explain the F150W flux. GL-z12 is now detected at higher significance in $F$150$W$ due to better overall handling of background features (e.g., wisps) in the SW filters, ruling out the secondary peak in the \texttt{Prospector} $p(z)$ distribution centered at $z\approx14$.

\end{CJK*}
\end{document}